\documentclass[twocolumn,showpacs,preprintnumbers,amsmath,amssymb]{revtex4}
\usepackage{longtable}
\usepackage{graphicx}% Include figure files
\usepackage{dcolumn}% Align table columns on decimal point
\usepackage{bm}% bold math
\usepackage{rotating}

\begin{document}

\preprint{submitted to JCP}
\title{Scalar and spin-dependent relativistic effects on magnetic properties calculated with four-component methods: the nuclear magnetic resonance parameters of the lead halides}
%\title{Fully relativistic {\em ab initio} calculation of nuclear magnetic resonance parameters of the lead dihalides}% Force line breaks with \\

\author{Rodolfo H. Romero}
%\altaffiliation[Present address: ]{Physics Department, Universidad Nacional del Nordeste, Avenida  Libertad 5500 (3400) Corrientes, Argentina.}%Lines break automatically or can be forced with \\
%\affiliation{ Chemistry Department, University of Helsinki, A.~I.~Virtasen aukio 1, FIN-00014, Finland.}
\affiliation{Physics Department, Facultad de Ciencias Exactas, Universidad Nacional del Nordeste, Avenida  Libertad 5500 (3400) Corrientes, Argentina.}
\email{rhromero@exa.unne.edu.ar}

%\affiliation{

\date{November 9, 2006}% It is always \today, today,

\begin{abstract}
The results of calculations of nuclear magnetic resonance (NMR) parameters for the lead halides is reported in this paper. The results are obtained by using four-component methods. The use of the nonrelativistic L\'evy-Leblond Hamiltonian along with the relativistic Dirac-Coulomb and spin-free ones allows us to discriminate scalar and spin-dependent effects on the parameters. It is found that the wide range of the lead NMR spectra and their large anisotropies are, mainly, due to  spin-dependent effects on the paramagnetic term. Among the relativistic scalar corrections, the so-called spin-Zeeman kinetic-energy term turns out to be dominant. The reduced spin-spin coupling constants become proportional to the product of the atomic numbers of the coupled nuclei.
\end{abstract}

\pacs{Valid PACS appear here}% PACS, the Physics and Astronomy
\maketitle

\section{Introduction}
It is nowadays well established that the calculation of properties of molecules containing heavy atoms requires the inclusion of the kinematic effects of relativity in order to correctly account for their magnitude and, in some cases, even their experimental trends \cite{Pyykko,Schwerdtfeger}. A number of methods and formalisms have been devised for such a purpose \cite{Schwerdtfeger, Aucar93, Quiney98, Visscher99,Wolff99, Melo03, Manninen03}; among them, {\em ab initio} fully relativistic four-component methods \cite{Aucar93, Quiney98, Visscher99} occupy a privileged position as a reference for benchmark calculations against which more approximated methods are to be compared. In particular, the calculation of the parameters of the nuclear magnetic resonance (NMR) spectroscopy are known to be highly sensitive to the relativistic effects \cite{Mason, Kaupp}. 

From an experimental point of view, NMR spectroscopy is a powerful tool for the determination of molecular structures, thus enabling its use for a wide range of applications. Among the magnetic nuclei, $^{207}$Pb, the heaviest one having spin $I=1/2$, is particularly suitable for discriminating among different structures and oxidation states in lead-containing compounds because of its natural abundance (22.6\%) and excellent receptivity \cite{Mason}. The high sensitivity of $^{207}$Pb to its coordination environment gives rise to large changes in its chemical shifts covering a range of about 16000 ppm for the known compounds. %The large polarizability of lead, due to its large number of electrons, has been pointed out as the cause for the broad $^{207}$Pb NMR spectra, reflected in large anisotropies in the chemical shift tensor.
Lead can form both inorganic and organolead compounds. Inorganic lead chemistry is dominated by divalent Pb(II) while organolead chemistry is based on tetravalent Pb(IV) \cite{Kaupp93}. Lead dihalides are among the simpler inorganic salts that divalent lead can form.
%It has been observed that group 14 halides systematically favor higher valence compounds for the lighter elements compared to the corresponding elements from the fifth and sixth rows for which oxidation state +4 is prefered \cite{Escalante99}.
%Thus, carbene analogues C$X_2$, where $X$ is a halogen, are highly reactive and Pb$X_4$ compounds are rare. The energy $\Delta U_g$, of disproportionation reactions, 2$MX_2$(g)$\rightarrow MX_4$(g)$+M$(g), have been used to analize the valence preference of some groups of the periodic table \cite{Escalante99, Seth98}.
PbF$_2$ is probably the most studied lead halide. Solid PbF$_2$ is a ionic solid presenting a strong superionic
conductivity. It exhibes two phases: an orthorombic $\alpha$-PbF$_2$ phase stable at high pressure and low
temperatures, and a cubic $\beta$-PbF$_2$ phase more stable in ambient conditions. \cite{Costales00}.
The absoption spectra of gaseous PbF$_2$ was first reported in Ref. \cite{Hauge68}. 
It was shown that molten PbF$_2$ vaporizes partly as molecular PbF$_2$ with a considerable desproportionation 
%66\% PbF$_4$, 25\% PbF$_2$, 2\% PbF and 7\% Pb at 715$^\circ$) 
\cite{Hauge68}.
Measurements of $^{207}$Pb chemical shifts in lead halide powders have been performed using the magic-angle
spinning (MAS) technique \cite{Dybowski98}. 
Calculations of $^{19}$F NMR shielding constants has been reported for a number of metal fluorides, including PbF$_2$ \cite{Cai02}.
Calculations of molecular structures and vibrational spectroscopic properties for the lead halides, among other group 14 compounds, have been performed 
%The results were obtained 
by using a correlated complete active space self-consistent field (CASSCF) and multireference singles-doubles configuration interaction (MRSDCI) wave function, and relativistic effective core potentials (ECP) \cite{Benavides94,Dai94}, 
%to take relativistic effects into account 
%The molecular structure of the group 14 dihalides was also the subject of the computational study of Ref.~\cite{Szabados03}, using 
the coupled cluster method with singles and doubles excitations and perturbative inclusion of triple excitations, CCSD(T), suplemented with ECPs \cite{Szabados03}, and 
%Escalante {\em et al} have also studied the energetics and molecular structure of group 14 di- and tetrahalides by using 
density functional theory (DFT) and ECPs \cite{Escalante99}.
To the author's knowledge, the only theoretical study of the NMR magnetic shieldings of isolated molecules of lead halides is the one presented in Ref. \cite{Nizam89}. In that paper it is also reported calculations in clusters that preserves the orthorombic symmetry of the solid and experimental measurements in solid samples \cite{Nizam89}. However, although an agreement is found with experimental trends, the basis sets used are small and the method of calculation involves too many approximations so that no precise quantitative conclusions can be drawn from the results.

In this paper we shall consider the fully relativistic calculation of the NMR parameters of the lead dihalides with a twofold purpose. 
Firstly, it is aimed to explain the wide-range variations of the NMR parameters, their large anisotropies, and their experimentally observed regularities as the different halogens are substituted, as well as to identify the underlying mechanisms beneath such a behavior. 
Secondly, the use of all-electron fully relativistic methods along with large-size basis sets, allows us to test the performance of these reference methods in systems of analogous molecules ranging from PbF$_2$, in which the central heavy atom is dominant, to PbI$_2$, constituted by three heavy atoms.
%%%%%%%%%%%%%%%%%%%%%%%%%%%%%%%%%%%%%%%%%%%%%%%%%%%%%%%%%%%%%%%%%%%%%%%%%%%%%%%%%%
%\section{Theory}
%%%%%%%%%%%%%%%%%%%%%%%%%%%%%%%%%%%%%%%%%%%%%%%%%%%%%%%%%%%%%%%%%%%%%%%%%%%%%%%%%%
\section{Calculation approach}
\subsection{Computational details}
Fully relativistic, spin-free and nonrelativistic calculations of the electronic structure of the lead dihalides Pb$X_2$ ($X=$ F, Cl, Br and I) were carried out self-consistently with the Dirac-Coulomb, spin-free \cite{Dyall94} and L\'evy-Leblond Hamiltonians \cite{Levy67}, respectively. 
The nuclear magnetic shielding tensor, $\sigma$, and the reduced spin-spin coupling tensor $K$ were calculated as four-component linear response functions using the polarization propagator approach within the Random Phase Approximation (RPA) \cite{Aucar93}. All calculations were performed with the {\sc dirac} code \cite{dirac}. It should be noted that even the nonrelativistic calculations were done with four-component methods by using the nonrelativistic L\'evy-Leblond Hamiltonian. Usually, fully relativistic and nonrelativistic calculations require different basis sets because of the different nature of the relativistic and nonrelativistic operators. All along the calculations reported here, the relativistic bases were used and have shown to provide convergence both in the spin-free and L\'evy-Leblond calculations. The use of a large value of the speed of light in fully relativistic calculations has also been advocated as a procedure to reach the nonrelativistic limit.
%; however, in some cases, the self-consistent calculation with such a procedure fails to converge unless a less accurate threshold is set.

The molecular structures were taken from experimental geometries \cite{Kaupp93,Hastie71}: $R(X{\rm -Pb})=2.033$, 2.46, 2.6 and 2.79 \AA,
and $\theta(X{\rm -Pb-}X)=97.8$, 96.0, 98.8 and 99.7$^\circ$ for $X=$ F, Cl, Br and I, respectively.
These bond lengths are close to values calculated with the MRSDCI method \cite{Benavides94}, but the calculated bond angles are about four degrees larger than the ones used here. So, there is some uncertainty inherent to our calculations because of their dependence on the molecular geometry.
In this work, large component uncontracted Cartesian Gaussian basis set were used in all atoms. 
The large component basis sets used in the calculations were obtained from modifying the Faegri basis sets (for Pb, Br and I) \cite{Faegri01} and the aug-ccpVTZ basis set (for F and Cl) by suplementing them with large- (tight functions) and small-exponent (diffuse functions) Gaussian basis functions. These tight (diffuse) functions were obtained from multiplying (dividing) by a factor of 3 the largest (smallest) exponent present in each block of angular momentum. The small component bases sets are generated from the large component bases by invoking the kinetic balance condition. 
The final ranges of the exponents in the expanded bases are similar to those used previously in accurate nearly basis set-limit calculations
%This procedure has proven to be lead to convergence on the basis set limit for the rare gas atoms 
\cite{Vaara03}.

%%%%%%%%%%%%%%%%%%%%%%%%%%%%%%%%%%%%%%%%%%%%%%%%%%%%%%%%%%%%%%%%%%%%%%%%%%%%%%%%%%
\subsection{Basis set optimization in PbBr$_2$.}
Given that the basis set completeness is a critical issue for reliable results, specially in the calculations of the two heaviest halides, the PbBr$_2$ molecule was taken as a model to test the basis set quality as a function of the functions added to the original basis sets for the large component. Table \ref{new PbBr2 convergence} shows the calculated values for the paramagnetic, diamagnetic and total nuclear magnetic shielding in both the lead and bromine nuclei and the one- and two-bonds indirect spin-spin coupling constants $^1J$($^{207}$Pb-$^{79}$Br) and $^2J$($^{79}$Br-$^{79}$Br). As one goes from basis A (Faegri basis set) to the large-size basis G, all the quantities converges steadily. In particular, the shielding diamagnetic terms $\sigma^{\rm d}$(Pb) and $\sigma^{\rm d}$(Br) and the two-bonds coupling $^2J$($^{79}$Br-$^{79}$Br) are reasonably well reproduced with the basis A already. In contrast, the  shielding paramagnetic terms and the one-bond coupling $^1J$($^{207}$Pb-$^{79}$Br) require of the large bases F or G to achieve convergence within about 10 ppm for $\sigma^{\rm p}$(Pb), less than 1 ppm for $\sigma^{\rm p}$(Br) and {\em ca.} 3 Hz for $^1J$(Pb-Br). Since no attempt was done to include electronic correlation effects, the basis F was considered converged enough and it was used for representing Pb and Br in all the calculations below. A similar basis scheme was adopted for the other halogens, including functions with exponents of the same order of magnitude as those used in similar atomic calculations \cite{Vaara03}. The effect of the tight, diffuse and polarization functions added was studied, although not shown here, to ensure that convergence was reached. 
With such criteria, the PbF$_2$, PbCl$_2$, PbBr$_2$ and PbI$_2$ molecules were described by 2168, 2288, 2332 and 2343 Gaussian basis functions, respectively.
The large and small component basis sets finally used are listed in Table \ref{basis}. 
%%%%%%%%%%%%%%%%%%%%%%%%%%%%%%%%%%%%%%%%%%%%%%%%%%%%%%%%%%%%%%%%%%%
\subsection{Neglect of $(SS|SS)$-integrals}
Following an approximation proposed by Visscher \cite{Visscher97}, the contribution from $(SS|SS)$-type two-electron integrals, {\em i.e.} those involving small components functions only, were neglected both in the wave function and the response calculations along this paper. This approximation allows a considerable saving in computational demands, typically decreasing the timings in one order of magnitud. However, an estimation of their importance was obtained by calculating their contributions in the PbBr$_2$ molecule with the smallest basis set A. The use of this basis set instead of the larger ones is justified because the small components of the molecular orbitals are densely centered in the atoms and only affect the electronic density in the very proximity of the nuclei, in contrast to the diffuse functions added to basis set A in the bases from B to G. Because of that, quite different combination of basis sets \cite{Faegri01,Hirao02} like Faegri(Pb)/Faegri(Br), Faegri(Pb)/Sadlej(Br) and Hirao(Pb)/Sadlej(Br) yield quite different $\sigma_{\rm iso}$(Pb) and $\sigma_{\rm iso}$(Br) far away from the converged values but, nevertheless, all of them give almost the same contributions from the inclusion of $(SS|SS)$-type two-electron integrals, namely, $\simeq$+50 ppm to $\sigma_{\rm iso}$(Pb) and {\em ca.} +1 ppm to $\sigma_{\rm iso}$(Br).
Hence, their contributions to $\sigma_{\rm iso}$(F) and $\sigma_{\rm iso}$(Cl) are negligible, while they contribute a few ppm to $\sigma_{\rm iso}$(I). All these contributions are below 1\% of the total values, thus justifying the exclusion of these integrals. Another approximation consisting in including the $(LL|SS)$-type integrals in the wave function calculation although neglecting them in the response calculation of the properties was also tried with similar results. It turns out that they typically contribute some $\simeq +40$ ppm to $\sigma_{\rm iso}$(Pb), somewhat less ($\simeq + 10$ ppm) for $\sigma_{\rm iso}$(I) and their effect is negligible for the lighter halogens. This approximation yields an extra saving in computing time and it could be recommendable in order to speed up computations without too much lost of accuracy. However, for consistency, all the numbers reported in this paper were obtained by fully including $(LL|SS)$ two-electron integrals in every step of the calculation.
%%%%%%%%%%%%%%%%%%%%%%%%%%%%%%%%%%%%%%%%%%%%%%%%%%%%%%%%%%%%%%%%%%%%%%%%%%%%%%%%%%
\begin{widetext}
\begin{table*}
%\begin{widetext}
\begin{ruledtabular}
\caption{\label{new PbBr2 convergence} Calculated values of the NMR parameters for PbBr$_2$ as a function of the basis set.}
\begin{tabular}{cllrrrrrrrr}
  &  \multicolumn{1}{c}{Large component}   &  & 
\multicolumn{3}{c}{$\sigma_{\rm iso}$(Pb)}& 
\multicolumn{3}{c}{$\sigma_{\rm iso}$(Br)}& & \\
\cline{4-6} \cline{7-9}
Basis  &  \multicolumn{1}{c}{functions}  &  No. &  param.& diam.& total & param.& diam. & total &  $^1J$(Pb-Br) &  $^2J$(Br-Br)  \\
\cline{1-3} \cline{4-6} \cline{7-9}  \cline{10-11}
         &  Pb: $23s\ 21p\ 14d\ 9f$ &  &         &         &          &         &         &         &          &         \\
     A   &  Br: $19s\ 16p\  9d$     & 1511 &-9466.4  & 7974.2  & -1492.2  & -216.2  & 2606.3  & 2390.1  & -3446.63 &  43.26  \\[5pt]
%-------------------------------------------------------------------------------------------------
         &  Pb: $27s\ 22p\ 17d\ 12f\ 2g\ 1h$ & & &   &   &    &    &             &            &        \\
     B   &  Br: $19s\ 16p\  9d\  1f$       & 1855 & -8119.8  &8025.1  & -94.7  & -498.5  & 2624.9  & 2126.4  &   -3847.67 &  46.77 \\[5pt]
%-------------------------------------------------------------------------------------------------
         &  Pb: $27s\ 22p\ 17d\ 12f\ 2g\ 1h$ &  &  &   &   &    &    &             &            &        \\
     C   &  Br: $19s\ 16p\  9d\  2f$       & 1905 & -8190.0  & 8026.1  & -163.1  & -529.8  & 2628.6  & 2098.8  &   -3866.52 &  46.18 \\[5pt]
%-------------------------------------------------------------------------------------------------
         &  Pb: $27s\ 22p\ 17d\ 12f\ 2g\ 1h$ & & &   &   &    &    &             &            &        \\
     D   &  Br: $19s\ 16p\  9d\  3f$       & 1955 & -8196.5  & 8027.4  & -169.9  & -530.3  & 2630.7  & 2100.4  &   -3868.19 &  46.13 \\[5pt]
%-------------------------------------------------------------------------------------------------
         & Pb: $27s\ 22p\ 17d\ 12f\ 2g\ 1h$  & & &   &   &    &    &             &            &        \\
     E   & Br: $20s\ 16p\ 13d\  2f\ 1g$     & 2125 & -8223.4  & 8028.0  & -195.3  & -545.3  & 2668.5  & 2123.2  &   -3891.24 &  46.16 \\[5pt]
%-------------------------------------------------------------------------------------------------
         & Pb: $27s\ 22p\ 17d\ 12f\ 3g\ 2h$ & & &   &   &    &    &             &            &        \\
     F   & Br: $20s\ 16p\ 13d\ 3f\ 2g$     & 2332 & -8122.9  & 8030.9  & -92.0  & -551.1  &2670.3  & 2119.2   &  -3950.06 & 44.99 \\[5pt]
%-------------------------------------------------------------------------------------------------
         & Pb: $27s\ 22p\ 17d\ 12f\ 4g\ 3h$  & & &   &   &    &    &             &            &        \\
     G   & Br: $20s\ 16p\ 13d\ 3f\ 2g$      & 2417 & -8113.9  & 8031.1  & -82.7  & -551.8  & 2670.6  & 2118.8  &   -3952.79 &  44.94 \\[5pt]
\end{tabular}
\end{ruledtabular}
%\end{widetext}
\end{table*}
\end{widetext}
%%%%%%%%%%%%%%%%%%%%%%%%%%%%%%%%%%%%%%%%%%%%%%%%%%%%%%%%%%%%%%%%%%%%%%%%%%%%%%%%%%

%%%%%%%%%%%%%%%%%%%%%%%%%%%%%%%%%%%%%%%%%%%%%%%%%%%%%%%%%%%%%%%%%%%
\begin{table}
\begin{ruledtabular}
\caption{\label{basis} Large and small component basis sets used in the calculations.}
\begin{tabular}{cllc}
Atom & Large component             & Small component                \\ \hline
F  & $13s\ 11p\ 9d\ 6f$            & $11s\ 22p\ 17d\ 9f\ 6g$        \\
Cl & $20s\ 13p\ 8d\ 7f$            & $13s\ 28p\ 20d\ 8f\ 7g$        \\
Br & $20s\ 16p\ 13d\ 3f\ 2g$       & $16s\ 22p\ 16d\ 13f\ 3g\ 2h$   \\
I  & $23s\ 20p\ 16d\ 2f\ 1g$       & $20s\ 28p\ 20d\ 16f\ 2g\ 1h$   \\
Pb & $27s\ 22p\ 17d\ 12f\ 3g\ 2h$  & $22s\ 29p\ 25d\ 17f\ 12g\ 3h\ 2i$ \\
\end{tabular}
\end{ruledtabular}
\end{table}
%%%%%%%%%%%%%%%%%%%%%%%%%%%%%%%%%%%%%%%%%%%%%%%%%%%%%%%%%%%%%%%%%%%
\section{Results and discussion}
\subsection{Absolute nuclear shieldings}
The results of nuclear magnetic shielding tensors for lead and halogen nuclei in the series of molecules PbF$_2$, PbCl$_2$, PbBr$_2$ and PbI$_2$ calculated with relativistic, spin-free and nonrelativistic Hamiltonians are shown in Table \ref{shield tensor}. The tensor ${\bm\sigma}$(Pb) is diagonal, with components $\sigma_\parallel$, $\sigma_{\perp}$ and $\sigma_z$ along the direction parallel to the molecular symmetry axis, perpendicular to it but in the plane of the molecule, and perpendicular to the molecular plane, respectively. In Table \ref{shield tensor} it is also shown the individual components of the paramagnetic and diamagnetic tensors ${\bm\sigma}^{\rm p}$ and ${\bm\sigma}^{\rm d}$, their corresponding isotropic values $\sigma_{\rm iso}=(1/3){\rm Tr}({\bm\sigma})$, and the span $\Omega=\sigma_1-\sigma_3$ defined as the diference between the largest and the smallest eigenvalues of the tensor ($\sigma_1 > \sigma_2 > \sigma_3$) for Pb and the halogen atoms.

Table \ref{SO-scalar} provides further insight on the results shown in Table \ref{shield tensor} by decomposing the various magnitudes in scalar and spin-dependent contributions given by
\begin{eqnarray}
\sigma^{\rm scalar} &=& \sigma^{\rm spin-free} - \sigma^{\rm NR}, \\
\sigma^{\rm SO} &=& \sigma^{\rm Rel} - \sigma^{\rm spin-free},
\end{eqnarray}
such that $\sigma^{\rm Rel} = \sigma^{\rm NR} + \sigma^{\rm scalar} + \sigma^{\rm SO}$. A similar decomposition for the spin-dependent terms has been recently applied to study of the static polarizability of group-13 atoms \cite{Fleig05}.
\subsubsection{Pb shielding}
\begin{figure}
\includegraphics[width=10cm]{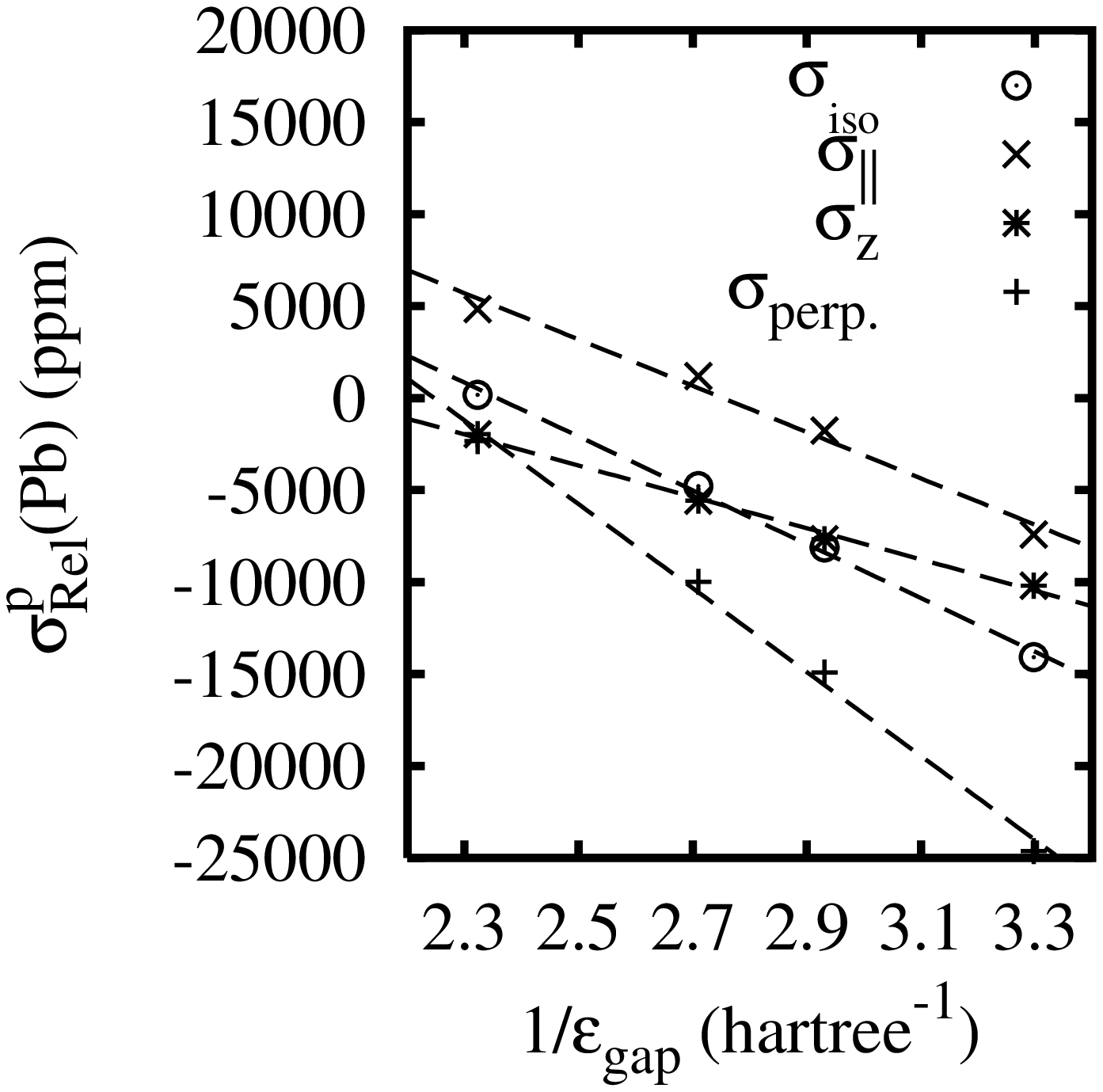}
\caption{\label{sigma-IP} Isotropic and tensor components of the paramagnetic contribution to the relativistic $^{207}$Pb absolute nuclear magnetic shielding in (from left to right) PbF$_2$, PbCl$_2$, PbBr$_2$ and PbI$_2$, as a function of the inverse of the HOMO-LUMO gap.}
\end{figure}

A clear trend to deshielding ($\sigma<0$) of the lead nucleus in Pb$X_2$ is observed from the relativistic (Rel), the spin-free (SF) and the nonrelativistic (NR) calculations as one goes from $X=$F to Cl to Br to I. However, this trend is much more pronounced for $\sigma_{\rm iso}^{\rm Rel}$(Pb) than for $\sigma_{\rm iso}^{\rm SF}$(Pb) or $\sigma_{\rm iso}^{\rm NR}$(Pb).
The analysis of $\sigma_{\rm iso}$ in terms of its paramagnetic and diamagnetic contributions shows that this is a consequence of the dramatic increase of the (negative) paramagnetic term from $\sim +200$ (in PbF$_2$) to $\sim -14000$ ppm (in PbI$_2$), that cannot be compensated by the slight growth in $\sigma_{\rm iso}^{\rm d}$ from 7896 to 8110 ppm. Furthermore, an inspection of the tensor components of the relativistic ${\bm\sigma}^{\rm p}$ and ${\bm\sigma}^{\rm d}$  shows that, to a large extent, ${\bm\sigma}^{\rm d}$ is almost isotropic, reflecting its deep core origin, while the different components of the tensor ${\bm\sigma}^{\rm p}$ behave distinctly in response to the kinematic effects of relativity, $\sigma_\perp^{\rm p}$ being more sensitive than $\sigma_z^{\rm p}$, and this, at its turn, more sensitive than $\sigma_\parallel^{\rm p}$, yielding $\sigma_\perp^{\rm p} < \sigma_z^{\rm p} < \sigma_\parallel^{\rm p}$ in all cases, both relativistically and nonrelativistically. It has been noted that the measured $^{207}$Pb isotropic shifts in the lead dihalides fit to a linear correlation with the inverse of the ionization potential \cite{Dybowski98}; such a correlation has been taken as an indication that the paramagnetic contribution to the chemical shift is dominant. Reciprocally, given that in our calculations the paramagnetic contributions are the most affected by relativistic effects, and taking the HOMO-LUMO gap as a measure of the ionization potential, Figure \ref{sigma-IP} shows that there is a linear correlation between each tensor component of the relativistically calculated $\sigma^{\rm p}$(Pb) with the inverse of $\varepsilon_{\rm HOMO-LUMO}$. The different slopes of the least-squares fitted straight lines can be taken as proportional to the corresponding matrix elements of the perturbations giving $\sigma^{\rm p}_\parallel$, $\sigma^{\rm p}_\perp$ and $\sigma^{\rm p}_z$.
\subsubsection{halogen atom shieldings}
%%%%%%%%%%%%%%%%%%%%%%%%%%%%%%%%%%%%%%%%%%%%%%%%%%%%%%%%%%%%%%%%%%%%%%%%%%%%%%%%%%
As shown in Table \ref{SO-scalar}, both $\sigma^{\rm p/SO}_{\rm iso}(X)$ and $\sigma^{\rm p/scalar}_{\rm iso}$($X$) becomes more positive as one changes the sustituent from F to I. The relative importance of their magnitude, however, changes from $|\sigma^{\rm p/SO}_{\rm iso}(X)| > |\sigma^{\rm p/scalar}_{\rm iso}(X)|$ for $X=$F and Cl, to the opposite for Br and I. 
There is no obvious interpretation for such a trend.
%It is not apparent the mechanism responsible for such a trend. 
In previous works, the study of the correlation of the various contributions to the nuclear charge $Z$ has been a useful guide for the identification of underlying mechanisms.
The best least-squares fits found for the paramagnetic shielding constants at the halogen nuclei is given by:
\begin{eqnarray}
%\sigma^{\rm p/scalar}_{\rm iso}(X) &=& 3.19126\times 10^{-3}Z^{3.28378}-27.777 \\
%\sigma^{\rm p/SO}_{\rm iso}(X)     &=& 1.11823\times 10^{-5}Z^{4.2867} -41.3122 
\sigma^{\rm p/scalar}_{\rm iso}(X) &=& 3.19\times 10^{-3}Z^{3.28}-27.8 \\
\sigma^{\rm p/SO}_{\rm iso}(X)     &=& 1.12\times 10^{-5}Z^{4.29} -41.3
\end{eqnarray}
Two facts deserve some attention: firstly, the different $Z$-dependence of those contributions, {\em i.e.}, $\sigma^{\rm p/scalar}_{\rm iso} \sim Z^3$ while $\sigma^{\rm p/SO}_{\rm iso} \sim Z^4$; and secondly, these correlations of the form $(AZ^n+B)$ fits better to the calculated values than the simpler ones $(AZ^n)$ which one would expect from a relativistic correction, in order to fulfill the requierement of vanishing when $Z$ goes to zero. No further insight can be provided within the present framework. A consideration of a perturbative treatment could be helpful in analyzing these trends more deeply.
%%%%%%%%%%%%%%%%%%%%%%%%%%%%%%%%%%%%%%%%%%%%%%%%%%%%%%%%%%%%%%%%%%%%%%%%%%%%%%%%%%
\begin{widetext}
\begin{table*}
\begin{ruledtabular}
\caption{\label{shield tensor} Isotropic values and tensor components of the nuclear magnetic shielding tensor of the lead and halogens nuclei $X$ ($X=$ F, Cl, Br, I) calculated with relativistic (Rel.), spin-free (SF) and nonrelativistic L\'evy-Leblond (NR) Hamiltonians. All values are given in ppm.
}
\begin{tabular}{lcrrrrrrrrrrrrrrr}
\multicolumn{1}{c}{Nucleus}  & \multicolumn{1}{c}{Quantity} &
\multicolumn{3}{c}{PbF$_2$}  & & \multicolumn{3}{c}{PbCl$_2$} & &
\multicolumn{3}{c}{PbBr$_2$} & &\multicolumn{3}{c}{PbI$_2$} \\ 
\cline{3-5} \cline{7-9} \cline{11-13} \cline{15-17} 
           &          &   Rel.    & SF&  NR   && Rel. & SF  &   NR   && Rel. & SF  &   NR   && Rel.  & SF  &   NR   \\ \hline
$^{207}$Pb &
 $\sigma_{\rm iso}$        &  8078.5   & 13226.5& 7465.4 &&  3152.0 & 11317.4 & 6101.3 &&  -92.0 & 10511.7 & 5616.1 && -5970.4 & 9273.7& 4763.8    \\
&$\sigma_{\rm iso}^{\rm p}$      &    182.2  & 5371.6& -2676.3 && -4790.8 & 3416.0 & -4086.7 && -8122.9 & 2519.7 &-4695.0 && -14080.7& 1202.8& -5651.4\\
&$\sigma_{\rm iso}^{\rm d}$      &   7896.4  & 7854.9& 10141.7 &&  7942.9 & 7901.4 & 10188.1 &&  8030.9 & 7992.0& 10311.1 &&  8110.2 & 8070.9& 10415.2 \\[5pt]
&$\sigma_\parallel$%\footnote{Tensor component in the direction of the symmetry axis of the molecule.}
                       &  12723.9  & 14233.1& 8561.4 &&  9120.4 & 12463.7 & 7463.3 &&  6239.6 & 11521.9 & 6954.7 &&  653.1 & 10105.2& 6125.8    \\
&$\sigma_z$%\footnote{Tensor component in the direction perpendicular to the plane of the molecule.}
                       &   5974.1  & 12721.5& 7083.4 &&  2427.1 & 11420.7 & 6216.2 &&   559.2 & 10982.3 & 6078.7 &&-1944.0 & 10640.8& 6041.3    \\
&$\sigma_\perp$%\footnote{Tensor component in the plane of the molecule, perpendicular to its symmetry axis.}
                       &   5537.5  & 12725.0& 6751.4 && -2091.4 & 10067.8& 4624.5 && -6974.9 & 9030.8 & 3815.0 && -16620.4& 7075.2& 2123.8  \\ & $\Omega$             &   7186.4  & 1508.1 & 1810.0 && 11211.8 & 2395.9 & 2838.8 && 13214.5 & 2491.1 & 3139.7 &&  17273.5& 3030.0 &4002.0 \\[5pt]
&$\sigma^{\rm p}_{\parallel}$&   4840.5  & 6391.3& -1567.0 &&  1199.5 & 4584.5 & -2702.4 && -1762.2 & 3558.8 & -3322.3 &&-7421.3 & 2070.1& -4245.6    \\
&$\sigma^{\rm p}_{z}$        &  -1963.7  & 4825.0& -3100.0 && -5578.1 & 3456.8 & -4034.7 && -7676.8 & 2884.6 & -4355.4 &&-10196.7& 2426.2& -4546.7    \\
&$\sigma^{\rm p}_{\perp}$    &  -2330.3  & 4898.6& -3361.8 && -9994.0 & 2206.7 & -5523.1 &&-14929.7 & 1115.7 & -6407.3 &&-24624.0& -887.8& -8161.8    \\[5pt]
&$\sigma^{\rm d}_{z}$        &   7937.9  & 7896.5& 10183.4 &&  8005.2 & 7963.9 & 10250.9 &&  8136.0 & 8097.6 & 10434.1 && 8252.7 & 8214.6& 10588.3    \\
&$\sigma^{\rm d}_{\parallel}$&   7883.4  & 7841.8& 10128.4 &&  7920.9 & 7879.2 & 10165.7 &&  8001.8 & 7963.1 & 10277.0 && 8074.4 & 8035.1& 10371.4    \\
&$\sigma^{\rm d}_{\perp}$    &   7867.8  & 7826.3& 10113.2 &&  7902.6 & 7861.1 & 10147.6 &&  7954.8 & 7915.2 & 10222.3 && 8003.6 & 7963.0& 10285.7    \\
\hline %%%%%%%%%%%%%%%%%%%%%%%%%%%%%%%%%%%%%%%%%%%%%%%%%%%%%%%%%%%%%%%%%%%%%%%
$X$%\footnote{$X=^{19}$F, $^{35}$Cl, $^{79}$Br}. 
&$\sigma_{\rm iso}$        &   220.3   &  263.1&  289.0 &&  641.4  &  678.6&  750.5&&  2119.2 & 2113.7 & 2260.1 &&  4453.7 & 4210.4&  3980.6 \\
&$\sigma_{\rm iso}^{\rm p}$      &  -261.7   & -219.0& -197.9 && -448.2  & -410.8& -415.2&&  -551.1 & -556.2 & -904.3 &&    91.8 & -142.4& -1580.4 \\
&$\sigma_{\rm iso}^{\rm d}$      &   482.0   &  482.1&  486.9 && 1089.6  & 1089.4& 1165.7&&  2670.3 & 2669.8 & 3164.4 &&  4361.9 & 4352.8&  5561.0 \\
&$\sigma_1$            &   458.9   &  462.0&  463.7 &&  994.1  &  993.2& 1023.6&&  2812.4 & 2810.1 & 2831.6 &&  5397.1 & 5357.5&  4812.6  \\
&$\sigma_2=\sigma_z$   &   219.3   &  268.0&  299.5 &&  766.4  &  794.9&  874.7&&  2494.7 & 2452.0 & 2613.7 &&  5355.9 & 4991.6&  4750.4  \\
&$\sigma_3$            &   -17.2   &   59.4&  103.9 &&  163.6  &  246.6&  353.3&&  1050.5 & 1078.9 & 1335.1 &&  2608.1 & 2282.1&  2378.7  \\
& $\Omega$             &   476.1   &  402.6&  359.8 &&  830.5  &  746.6&  670.3&&  1761.9 & 1731.2 & 1496.5 &&  2789.0 & 3075.4&  2433.9\\ [5pt]
%--------------------------------------------------------------------------------------------------------------------------------------
&$\sigma^{\rm p}_{1}$        &   -50.6   &  -48.5&  -53.0 && -118.6  & -120.5& -168.8&&   116.3 &  115.6 & -364.9 &&  1008.7&  980.6&  -784.7    \\
&$\sigma^{\rm p}_{z}$        &  -258.8   & -210.3& -183.0 && -327.2  & -298.7& -294.6&&  -187.7 & -230.1 & -565.3 &&   974.2&  618.6&  -834.1    \\
&$\sigma^{\rm p}_{3}$        &  -475.5   & -398.2& -357.6 && -899.0  & -815.0& -782.3&& -1581.9 &-1554.0 &-1782.8 && -1707.4&-2026.4& -3122.4    \\[5pt]
&$\sigma^{\rm d}_{1}$        &   509.6   &  509.7&  515.5 && 1112.7  & 1112.7& 1189.7&&  2696.3 & 2696.1 & 3194.6 &&  4390.7& 4382.1&  5597.2    \\
&$\sigma^{\rm d}_{z}$        &   478.1   &  478.2&  482.5 && 1093.6  & 1093.6& 1169.2&&  2682.4 & 2682.1 & 3178.9 &&  4381.6& 4373.0&  5584.5    \\
&$\sigma^{\rm d}_{3}$        &   458.3   &  458.5&  462.7 && 1062.6  & 1062.6& 1138.2&&  2632.2 & 2631.3 & 3119.8 &&  4313.2& 4303.2&  5501.2    \\
\end{tabular}
\end{ruledtabular}
%\flushleft{
%\begin{small}
% $^a$Tensor component in the direction of the symmetry axis of the molecule.\\
% $^b$Tensor component in the direction perpendicular to the plane of the molecule.\\
% $^c$Tensor component in the plane of the molecule, perpendicular to its symmetry axis.\\
% $^{\rm d}X=$F, Cl, Br, I.
% \end{small}
% }
\end{table*}
\end{widetext}
%%%%%%%%%%%%%%%%%%%%%%%%%%%%%%%%%%%%%%%%%%%%%%%%%%%%%%%%%%%%%%%%%%%%%%%%%%%%%%%%%%
\begin{widetext}
\begin{table*}
\begin{ruledtabular}
\caption{\label{SO-scalar} Isotropic values and tensor components of the spin-orbit and scalar contributions to the nuclear magnetic shielding tensor of the lead and halogens nuclei. All values are given in ppm.
}
\begin{tabular}{lcrrrrrrrrrrr}
\multicolumn{1}{c}{Nucleus}  & \multicolumn{1}{c}{Quantity} &
\multicolumn{2}{c}{PbF$_2$}  & & \multicolumn{2}{c}{PbCl$_2$} & &
\multicolumn{2}{c}{PbBr$_2$} & &\multicolumn{2}{c}{PbI$_2$} \\ 
\cline{3-4} \cline{6-7} \cline{9-10} \cline{12-13} 
           &          &   SO    & Scalar && SO & Scalar && SO & Scalar && SO  & Scalar  \\ \hline
$^{207}$Pb &
 $\sigma_{\rm iso}$        &  -5148.0 &   5761.1 && -8165.4 &  5216.1 && -10603.7 &  4895.6 &&-15244.1 &  4509.9 \\
 &$\sigma_{\rm iso}^{\rm p}$     &  -5189.4 &   8047.9 && -8206.8 &  7502.7 && -10642.6 &  7214.7 &&-15283.5 &  6854.2 \\
&$\sigma_{\rm iso}^{\rm d}$      &     41.5 &  -2286.8 &&    41.5 & -2286.7 &&     38.9 & -2319.1 &&    39.3 & -2344.3 \\[5pt]
%--------------------------------------------------------------------------------------------
&$\sigma_\parallel$    &  -1509.2 &   5671.7 && -3343.3 &  5000.4 &&  -5282.3 &  4567.2 && -9452.1 &  3979.4 \\
&$\sigma_z$            &  -6747.4 &   5638.1 && -8993.6 &  5204.5 && -10423.1 &  4903.6 &&-12584.8 &  4599.5 \\
&$\sigma_\perp$        &  -7187.5 &   5973.6 &&-12159.2 &  5443.3 && -16005.7 &  5215.8 &&-23695.6 &  4951.4 \\
& $\Omega$             &        &        &&       &       &&        &       &&       &       \\[5pt]
&$\sigma^{\rm p}_{\parallel}$&  -1550.8 &   7958.3 && -3385.0 &  7286.9 &&  -5321.0 &  6881.1 && -9491.4 &  6315.7 \\
&$\sigma^{\rm p}_{z}$        &  -6788.7 &   7925.0 && -9034.9 &  7491.5 && -10561.4 &  7240.0 &&-12622.9 &  6972.9 \\
&$\sigma^{\rm p}_{\perp}$    &  -7228.9 &   8260.4 &&-12200.7 &  7729.8 && -16045.4 &  7523.0 &&-23736.3 &  7274.0 \\[5pt]
%--------------------------------------------------------------------------------------------
&$\sigma^{\rm d}_{z}$        &   41.4 & -2286.9&&  41.3 &-2287.0&&   38.4 &-2336.5&&  38.1 &-2373.7\\
&$\sigma^{\rm d}_{\parallel}$&   41.6 & -2286.6&&  41.7 &-2286.5&&   38.7 &-2313.9&&  39.3 &-2336.3\\
&$\sigma^{\rm d}_{\perp}$    &   41.5 & -2286.9&&  41.6 &-2286.6&&   39.6 &-2307.1&&  40.6 &-2322.7\\
\hline %%%%%%%%%%%%%%%%%%%%%%%%%%%%%%%%%%%%%%%%%%%%%%%%%%%%%%%%%%%%%%%%%%%%%%%
$X^{\rm d}$%\footnote{$X=^{19}$F, $^{35}$Cl, $^{79}$Br}. 
&$\sigma_{\rm iso}$        &  -42.8 &  -25.9 &&-37.2& -71.9 &&   5.5  & -146.4&& 243.3 &  229.8\\
&$\sigma_{\rm iso}^{\rm p}$      &  -42.8 &  -21.1 &&-37.4&   4.4 &&   5.0  &  348.1&& 234.2 & 1438.0\\
&$\sigma_{\rm iso}^{\rm d}$      &   -0.1 &   -4.8 &&  0.2& -76.3 &&   0.5  & -494.6&&   9.1 &-1208.2\\[5pt]
&$\sigma_1$            &   -3.1 &   -1.7 &&  0.9& -30.4 &&   2.3  &  -21.5&&  39.6 &  544.9\\
&$\sigma_2=\sigma_z$   &  -48.7 &  -31.5 &&-28.5& -79.8 &&  42.7  & -161.7&& 364.3 &  241.2\\
&$\sigma_3$            &  -76.6 &  -44.5 &&-83.0&-106.7 && -28.4  & -256.2&& 326.0 &  -96.6\\
& $\Omega$             &   73.5 &   42.8 && 83.9&  76.3 &&  30.7  &  234.7&&-286.4 &  641.5\\[5pt]
%--------------------------------------------------------------------------------------------------------------------------------------
&$\sigma^{\rm p}_{1}$        &   -2.1 &    4.5 &&  1.9&  48.3 &&   0.7  &  462.5&&  28.1 & 1765.3\\
&$\sigma^{\rm p}_{z}$        &  -48.5 &  -27.3 &&-28.5&  -4.1 &&  42.4  &  335.2&& 355.6 & 1452.7\\
&$\sigma^{\rm p}_{3}$        &  -77.3 &  -40.6 &&-84.0& -32.7 && -27.9  &  228.8&& 319.0 & 1096.0\\[5pt]
&$\sigma^{\rm d}_{1}$        &   -0.1 &   -5.8 &&  0.0& -77.0 &&   0.2  & -498.5&&   8.6 &-1215.1\\
&$\sigma^{\rm d}_{z}$        &   -0.1 &   -4.3 &&  0.0& -75.6 &&   0.3  & -496.8&&   8.6 &-1211.5\\
&$\sigma^{\rm d}_{3}$        &   -0.2 &   -4.2 &&  0.0& -75.6 &&   0.9  & -488.5&&  10.0 &-1198.0\\
\end{tabular}
\end{ruledtabular}
\end{table*}
\end{widetext}
%%%%%%%%%%%%%%%%%%%%%%%%%%%%%%%%%%%%%%%%%%%%%%%%%%%%%%%%%%%%%%%%%%%%%%%%%%%%%%%%%%
%\begin{widetext}
%%%%%%%%%%%%%%%%%%%%%%%%%%%%%%%%%%%%%%%%%%%%%%%%%%%%%%%%%%%%%%%%%%%%%%%%%%%%%%%%%%%
\subsection{$^{207}$Pb shifts}
The standard reference for $^{207}$Pb chemical shifts measurements is tetramethyllead (TML), Pb(CH$_3$)$_4$, {\em i.e.}, the chemical shifts are defined as $\delta$($^{207}$Pb)$=\sigma({\rm TML})-\sigma$(Pb$X_2$). Nevertheless, we shall estimate here the lead chemical shifts by refering our compounds to plumbane (PbH$_4$), {\em i.e.}, with respect to a reference in which every methyl group was replaced by a hydrogen atom; such a change should not introduce severe changes while keeping the computational costs withing reasonably bounds. 
The calculation of plumbane NMR parameters is not only useful for obtaining the measurable shifts but also because of its tetrahedral symmetry $T_d$, what provides an interesting example for the analysis of the spin-dependent and scalar contributions to the nuclear magnetic shielding tensor of lead.
Because of the $T_d$ symmetry, the tensor $\sigma$(Pb) as well as its paramagnetic and diamagnetic contributions become isotropic.
The results of Dirac-Coulomb, spin-free and nonrelativistic calculations are shown in Table \ref{PbH4}. It can be seen that the spin-dependent contributions to both the paramagnetic and diamagnetic term represent a small fraction of the total value, while almost the total relativistic corrections are due to scalar effects. Furthermore, the magnitude of those effects ($\sigma^{\rm p/scalar}_{\rm iso}\simeq 8000$ ppm, $\sigma^{\rm d/scalar}_{\rm iso}\simeq -2300$ ppm) are quite close to, {\em e. g.}, those in PbF$_2$, thus showing that they arise from the inner most electrons in the core.
The isotropic lead chemical shifts and its paramagnetic and diamagnetic contributions are also shown in Table \ref{PbH4}. The decomposition of the relativistic corrections to the shifts ($\delta^{\rm Rel}-\delta^{\rm NR}$) in terms of scalar ($\delta^{\rm scalar}$) and spin-dependent term $\delta^{\rm SO}$ shows that almost the entire correction is due to the spin-orbit contribution to the paramagnetic term $\delta^{\rm p/SO}$ in contrast to $\delta^{\rm d/SO}$ which becomes negligible because $\delta^{\rm d}$ comes mainly from contributions from the spherically symmetric inner most atomic-like $s$ orbitals. The magnitude of the contributions becomes $\delta^{\rm d/SO} < \delta^{\rm d/scalar} < \delta^{\rm p/scalar} < \delta^{\rm p/SO}$, with $\delta^{\rm p/SO}$ accounting for more than 90\% of $\delta_{\rm iso}$ even for PbI$_2$.
\begin{table}
\begin{ruledtabular}
\caption{\label{PbH4} Isotropic nuclear magnetic shielding constant of lead nucleus in plumbane calculated with relativistic (Rel.), spin-free (SF) and nonrelativistic L\'evy-Leblond (NR) Hamiltonians, and spin-orbit (SO) and scalar contributions to them, and isotropic chemical shifts of the Pb nuclei in the lead halides with respect to plumbane, taken as reference compound. All values are given in ppm.
}
\begin{tabular}{ccrrrrr}
Molecule &                  &   Rel.   &   SF    &   NR    &   SO  &   scalar \\[5pt] \hline
PbH$_4$ & $\sigma$    & 12776.7  & 12674.2 &  6869.5 & 102.5 &  5804.7  \\
        & $\sigma^{\rm p}$  &  4941.3  &  4880.4 & -3211.5 &  60.9 &  8091.9  \\
        & $\sigma^{\rm d}$  &  7835.4  &  7793.8 & 10081.0 &  41.6 & -2287.2  \\
\hline %---------------------------------------------------------------------------
PbF$_2$ & $\delta$    & 4698.2   &  -552.3 &  -595.9 & 5250.5&    43.6   \\
        & $\delta^{\rm p}$  & 4759.1   &  -491.2 &  -535.2 & 5250.4&    44.0   \\
        & $\delta^{\rm d}$  &  -61.0   &   -61.1 &   -60.7 &    0.1&    -0.4   \\[5pt]
%----------------------------------------------------------------------------------
PbCl$_2$& $\delta$    & 9624.7   &  1356.8 &   768.2 & 8267.9&   588.6   \\
        & $\delta^{\rm p}$  & 9732.1   &  1464.4 &   875.2 & 8267.7&   589.2   \\
        & $\delta^{\rm d}$  & -107.5   &  -107.6 &  -107.1 &    0.1&     0.5   \\[5pt]
%----------------------------------------------------------------------------------
PbBr$_2$& $\delta$    & 12868.7  &  2162.5 &  1253.4 & 10706.2&  909.1    \\
        & $\delta^{\rm p}$  & 13064.2  &  2360.7 &  1483.5 & 10703.5&  877.2    \\
        & $\delta^{\rm d}$  &  -195.5  &  -198.2 &  -230.1 &     2.7&   31.9    \\[5pt]
%----------------------------------------------------------------------------------
PbI$_2$& $\delta$     & 18747.1  &  3400.5 &  2105.7 & 15346.6& 1294.8    \\
        & $\delta^{\rm p}$  & 19022.0  &  3677.6 &  2439.9 & 15344.4& 1237.7    \\
        & $\delta^{\rm d}$  &  -274.8  &  -277.1 &  -334.2 &     2.3&   57.1    \\[5pt]
\end{tabular}
\end{ruledtabular}
\end{table}
%%%%%%%%%%%%%%%%%%%%%%%%%%%%%%%%%%%%%%%%%%%%%%%%%%%%%%%%%%%%%%%%%%%%%%%%%%%%%%%%%%%
\subsection{Reduced spin-spin couplings}
The inclusion of relativistic effects in the calculation of indirect spin-spin coupling constants of lead-containing compounds has been considered through a number of methods, {\em e.g.}, semiempirically \cite{Pyykko81,Gonzalez97}, DFT-ZORA approximation \cite{Bryce02, Autschbauch00} and fully relativistic polarization propagator approach \cite{Enevoldsen00}, although, to the author's knowledge, there are no calculations for the couplings in the lead dihalides.
In Table \ref{spin-spin}, the results of nonrelativistic (NR) and relativistic (Rel) calculations are shown along with the scalar and spin-dependent (SO) contributions. As with the $\sigma$ and $\delta$ calculations, $K^{\rm scalar}$ and $K^{\rm SO}$ are obtained as a difference with respect to the result of a spin-free calculation. It can be seen that both isotropic and anisotropic one-bond couplings becomes more and more negative as one goes from the fuoride to the iodide. Both the scalar and spin-dependent contributions give rise to such a trend, with the scalar mechanisms being dominant.

\begin{figure}
\begin{tabular}{cc}
\includegraphics[width=9cm]{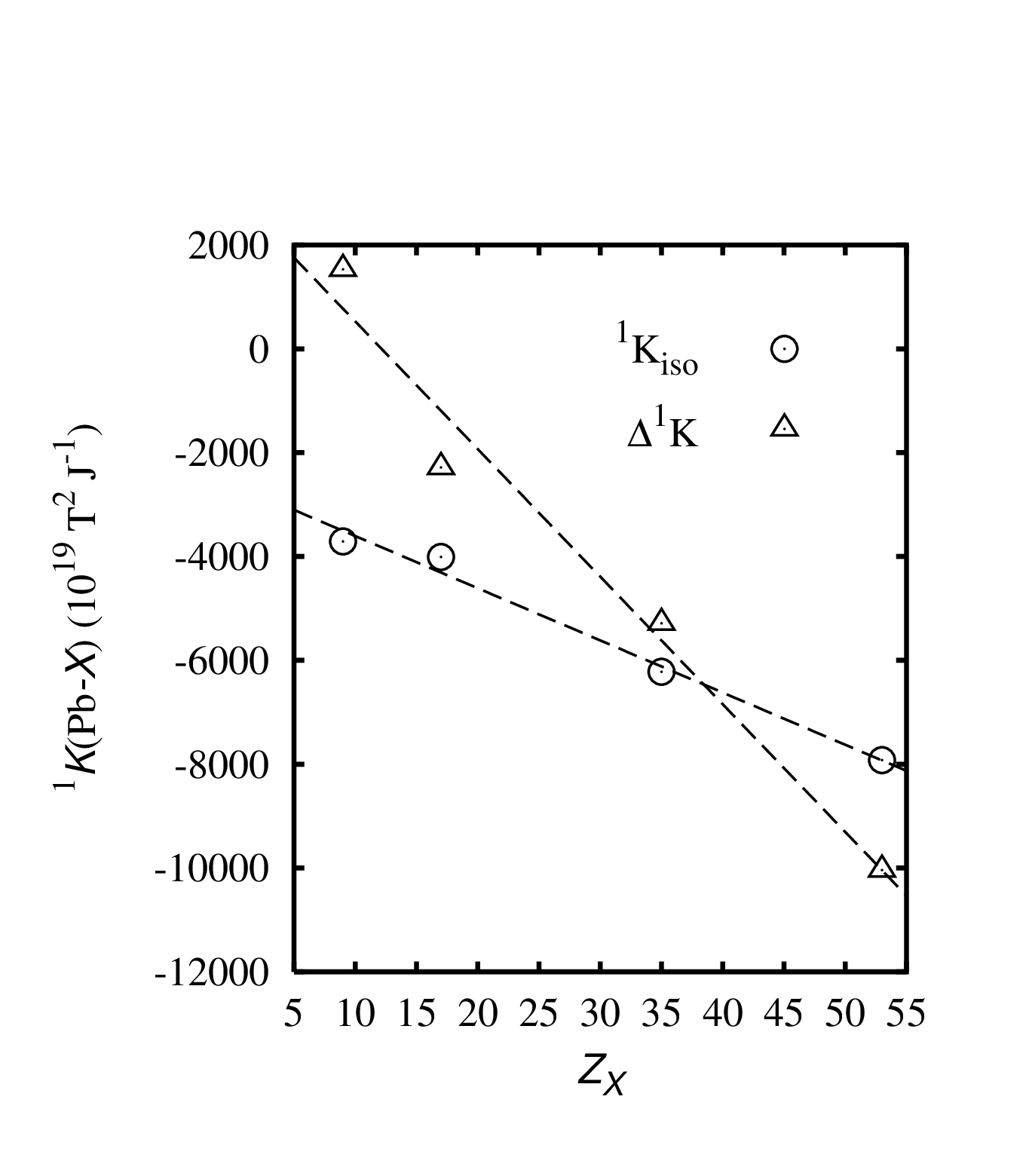} &
\includegraphics[width=9cm]{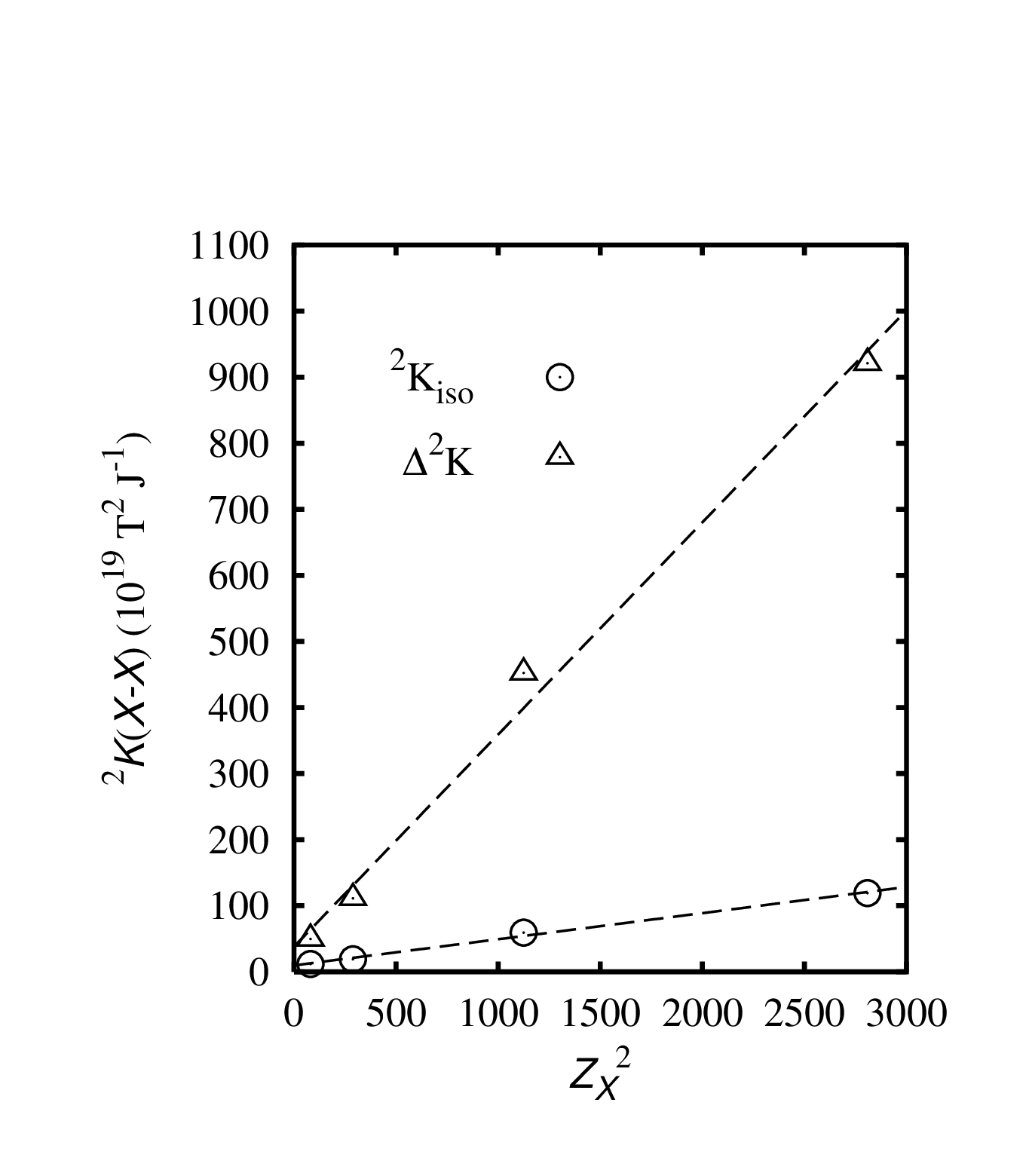} 
\end{tabular}
\caption{\label{plot K-Z} Relativistic one- and two-bonds reduced isotropic ($K_{\rm iso}$) and anisotropic ($\Delta K$) spin-spin coupling constants as a function of the halogen nuclear charge $Z_X$.}
\end{figure}

On the contrary, two-bonds couplings show a decreasing behavior in nonrelativistic calculations, as opposite to the increasing trend of the relativistic ones, as one goes from F to Cl to Br to I. This is a consequence of, mainly, the SO and --to less extent-- scalar mechanisms, both of which give positive contributions to $^2K_{\rm iso}$ and $\Delta^2K$. Nevertheless, it is worthy to note that, for the heaviest halide, PbI$_2$, $\Delta^2K^{\rm scalar}$(I$-$I) becomes greater than $\Delta^2K^{\rm SO}$(I$-$I).
There have been attempts in the literature to correlate the reduced spin-spin coupling constants with the square of the atomic number of a central heavy atom bonded to light atoms. On the other hand, Bryce {\em et al.} found that $K_{\rm iso}$ and $\Delta K$ have a good linear correlation with the product of atomic numbers of the coupled nuclei in interhalogen diatomics \cite{Bryce02}. In the Figure \ref{plot K-Z} we plot relativistically calculated $^1K_{\rm iso}$ and $\Delta^1K$ as a function of the halogen atomic number $Z_X$ (left panel) and $^2K_{\rm iso}$ and $\Delta^2K$ as a function of $Z^2_X$ (right panel panel). The dashed line represents the least-squares fit of the data. As it can be seen, the correlation is quite reasonable, especially for the isotropic values and somewhat less for the anisotropies. These results provide support to the correlation proposed in Ref. \cite{Bryce02}, generalizing it to include two-bonds couplings.
%%%%%%%%%%%%%%%%%%%%%%%%%%%%%%%%%%%%%%%%%%%%%%%%%%%%%%%%%%%%%%%%%%%%%%%%%%%%%%%%%%
\begin{table}
\begin{ruledtabular}
\caption{\label{spin-spin} Isotropic ($K_{\rm iso}$) and anisotropic ($\Delta K$) reduced one- and two-bonds spin-spin coupling constants of the lead halides and their scalar and spin-dependent contributions calculated with nonrelativistic L\'evy-Leblond (NR) and relativistic spin-free and Dirac-Coulomb (Rel.) Hamiltonians. All values are given in 10$^{19}{\rm T}^2\cdot{\rm J}^{-1}$.
}
\begin{tabular}{ccrrrrr}
& & \multicolumn{2}{c}{Pb-$X$} && \multicolumn{2}{c}{$X$-$X$} \\
\cline{3-4} \cline{6-7}
Molecule & Contribution\footnote{$K^{\rm Rel}=K^{\rm NR}+K^{\rm scalar}+K^{\rm SO}$ with $K^{\rm scalar}=K^{\rm spin-free}-K^{\rm NR}$, and  $K^{\rm SO}=K^{\rm Rel}-K^{\rm spin-free}$.} &
                  $^1K_{\rm iso}$ & $\Delta K$ &&$^2K_{\rm iso}$ & $\Delta K$   \\ \hline 
PbF$_2$  &  NR    &  -652.02    &   504.44   &&      5.21  &     13.86  \\
         &  scalar& -2397.07    & -1236.65   &&      1.01  &     10.00  \\
         &  SO    &  -661.17    &  2262.58   &&      5.62  &     25.72  \\
	 &  Rel.  & -3710.26    &  1530.37   &&     11.84  &     49.58  \\[5pt]
%--------------------------------------------------------------------------
PbCl$_2$ &  NR    &  -718.39    &   536.92   &&     -2.09  &     47.94  \\
         &  scalar& -2427.97    & -1980.39   &&      7.02  &     10.72  \\
         &  SO    &  -863.33    &  -841.05   &&     14.21  &     52.57  \\
	 &  Rel.  & -4009.69    & -2284.52   &&     19.14  &    111.23  \\[5pt]
%--------------------------------------------------------------------------
PbBr$_2$ &  NR    & -1261.10    &  -991.00   &&    -22.74  &    218.95  \\
         &  scalar& -3384.23    & -3011.64   &&     23.20  &     67.45  \\
         &  SO    & -1573.86    & -1282.44   &&     58.80  &    165.90  \\
	 &  Rel.  & -6219.19    & -5285.08   &&     59.26  &    452.30  \\[5pt]
%--------------------------------------------------------------------------
PbI$_2$  &  NR    & -1778.76    &  -1521.95  &&    -83.71  &    375.45  \\
         &  scalar& -4101.85    &  -6170.91  &&     26.72  &    326.95  \\
         &  SO    & -2042.77    &  -2348.63  &&    176.05  &    218.92  \\
	 &  Rel.  & -7923.38    & -10041.49  &&    119.06  &    921.32  \\
\end{tabular}
\end{ruledtabular}
\end{table}
%%%%%%%%%%%%%%%%%%%%%%%%%%%%%%%%%%%%%%%%%%%%%%%%%%%%%%%%%%%%%%%%%%%%%%%%%%%%%%%%%%%%%%%
\begin{widetext}
\end{widetext}
%%%%%%%%%%%%%%%%%%%%%%%%%%%%%%%%%%%%%%%%%%%%%%%%%%%%%%%%%%%%%%%%%%%%%%%%%%%%%%%%%%
%%%%%%%%%%%%%%%%%%%%%%%%%%%%%%%%%%%%%%%%%%%%%%%%%%%%%%%%%%%%%%%%%%%%%%%%%%%%%%%%%%%									 
\section{Conclusions}
In summary, in this paper we have reported results of calculations of the NMR parameters for the lead halides. The calculations made use of four-component relativistic Dirac-Coulomb and nonrelativistic L\'evy-Leblond Hamiltonians. In addition, the use of the spin-free hamiltonian allows to separate the scalar and spin-dependent relativistic contributions. 

The results presented in this paper can be rationalized as follows: 

{\em (i)} the main relativistic correction to the nuclear magnetic constants in lead is the spin-dependent contribution $\sigma^{\rm p/SO}$(Pb) which can be roughly associated with the spin-orbit interaction, strongly affected by the valence structure. Substitution of increasingly heavy halogens changes dramatically such a structure from PbF$_2$ dominated by the lead nucleus to PbI$_2$, a three heavy atoms molecule. 

{\em (ii)} $\sigma^{\rm p/scalar}$(Pb) ranges from 8047 ppm for PbF$_2$ to 6854 ppm for PbI$_2$, what is compatible with a dominant $\sigma^{\rm SZ-KE}$ term estimated in 7414 ppm by using an analytical model for the shielding in atoms \cite{Romero06}. However, the corresponding estimations for F (9 ppm) and Cl (64 ppm) are not good because of the large anisotropies of their $\sigma^{\rm p/scalar}$ tensors; the model estimates 570 and 1995 ppm for Br and I, in moderate agreement with the values of Table \ref{SO-scalar} (348 and 1438 ppm). The reason for that is that the spherical symmetry assumed in the model is more approximately realized in the heaviest halogen nuclei.

{\em (iii)} The $\sigma^{\rm d/scalar}$ term was studied in Ref. \cite{Gomez05}, where it has been shown that, within a perturbative framework, it arises from two linear response contributions (Darwin and mass-velocity) plus three expectation value terms, with all of them following a $Z^3$ power-law scaling with the atomic number; {\em e.g.}, the perturbative LR-ESC method gives $\sigma^{\rm d/scalar}({\rm Rn})=-3446.4$ ppm \cite{Gomez05}, whose extrapolation to Pb would give $\sigma^{\rm d/scalar}({\rm Pb})=-3446.4\times(82/86)^3 = -2987.5$ ppm, in line with the value of about -2300 ppm obtained for all halides. Performing the same extrapolation from the LR-ESC value for the nearest rare gas atom (taken from Ref.\cite{Gomez05}) to each halogen, one has -3.3, -23.8, -215.1 and -767.1 ppm, to be compared to -4.8, -76.3, -494.6 and -1208.2 ppm of the present work (Table \ref{SO-scalar}). Although the trend is correct, a lack of quantitative agreement still remains.

{\em (iv)} The $\sigma^{\rm d/SO}$ term can be disregarded in all cases, what can be attributed to the negligible effect of the spin-orbit interaction upon the inner electronic shells, which are responsible for the diamagnetic shielding. \\

The large spin-dependent effects on the paramagnetic shielding are also responsible for the large observed variations of the lead resonance over a wide range, as well as its large anisotropies, which originates in the different responses of the molecule to the spin-orbit interactions, although every component become proportional to the inverse of the HOMO-LUMO gap. The same trends are found in the $^{207}$Pb chemical shifts in the various halides because of the isotropic nature of the relativistic corrections, having a scalar origin, in the reference compound (plumbane).\\

Both the scalar and the spin-dependent mechanisms are important for the isotropic reduced coupling constants $^1K_{\rm iso}$ and $^2K_{\rm iso}$, although $|^1K_{\rm iso}^{\rm scalar}|>|^1K_{\rm iso}^{\rm SO}|$, while $|^2K_{\rm iso}^{\rm scalar}|<|^2K_{\rm iso}^{\rm SO}|$. Finally, the relativistic one- and two-bonds reduced coupling constants becomes well correlated to the product of the nuclear charges of the coupled nuclei, $Z_{\rm Pb}Z_X$ and $Z_X^2$, respectively.\\
\section*{Acknowledgements}
I thank T.~Saue (Strasbourg) for providing a pre-release version of the {\sc dirac} code during the early stages of this work.\\
This work was partly supported by SECyT (Universidad Nacional del Nordeste) and CONICET (Argentina).
%##################################################################################

\end{document}